# Remark on "Algorithm 680: evaluation of the complex error function": Cause and Remedy for the Loss of Accuracy Near the Real Axis

MOFREH R. ZAGHLOUL, United Arab Emirates University[1]

In this remark we identify the cause of the loss of accuracy in the computation of the Faddeyeva function, $w(z)$, near the real axis when using Algorithm 680. We provide a simple correction to this problem which allows us to restore this code as one of the important reference routines for accuracy comparisons.

---

Author's addresses: M. Zaghloul, Department of Physics, College of Sciences, United Arab Emirates University, Al-Ain, 15551, UAE.

## 1. INTRODUCTION

By modifying the tuning of Algorithm 363 [Gautschi 69, 70], using a different approximation near the origin and testing the relative rather than the absolute error, Algorithm 680 [Poppe and Wijers. 1990a,b] calculates the Faddeyeva function to 14 significant digits (in the first quadrant) with a significant increase in speed over Algorithm 363. The major modification made was in the choice of the contour $\Gamma$ used to tune the algorithm from the region R to the outer region Q, as depicted in Fig. 1, where the *v*th convergent of the Laplace continued fraction is used to approximate the Faddeyeva function asymptotically to a prescribed accuracy of *d* significant digits.

Figure 1: Regions and contours used in Algorithm 680

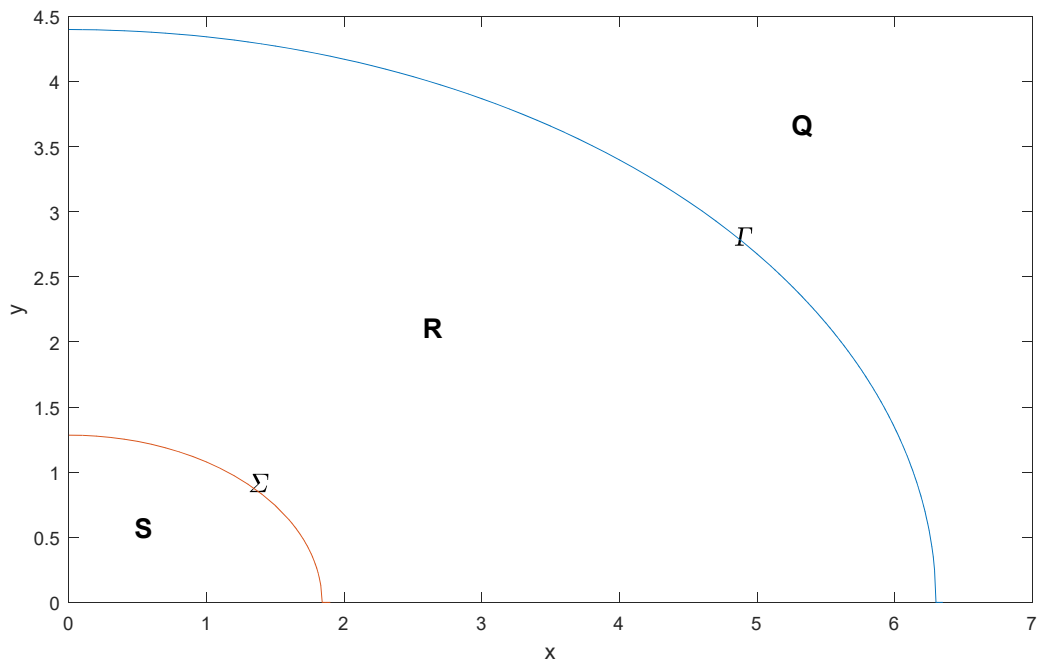

The contour $\Gamma$ in Algorithm 680 was defined by the condition

$$\rho(z) = \sqrt{\left(\frac{x}{x_0}\right)^2 + \left(\frac{y}{y_0}\right)^2} = 1 \qquad (1),$$

where $z=x+iy$ is the complex argument of the Faddeyeva function $w(z)$, $x_0$=6.3 and $y_0$=4.4. In the region Q between 3 and 16 convergents of the continued fraction are used to approximate the function to obtain a claimed accuracy of 14 significant figures. In the region R, the Faddeyeva function is evaluated by a truncated "downward" Taylor expansion where the Laplace continued fraction is used to calculate the derivatives of $w(z)$. In the innermost region,

S, constrained by the contour $\Sigma$ defined by $0 \leq \rho(z) \leq 0.292$, the function is evaluated using a power-series [Abramowitz and Stegun (7.1.5)].

## 2. Cause and Remedy

As pointed out by [Zaghloul and Ali 2011; Fig. 1(a)] Algorithm 680 loses its accuracy near the real axis. This loss of accuracy is in the regions R and Q and is most prominent in the neighborhood of the tuning contour $\Gamma$ or around $x=6.3$. In a recent work [Zaghloul 2018], the region of application for the asymptotic approximation of the Faddeyeva function using Laplace continued fraction has been assessed through a systematic comparison with Algorithm 916 as a reference. For a targeted accuracy of 13 significant-digits, the region of applicability of the asymptotic approximation from Laplace continued fraction, for very small $y$, is found to be $x \geq 20$. This explains the cause behind the loss of accuracy of Algorithm 680 near the real axis where the Laplace continued fraction is used directly down to $x=6.3$ and indirectly in region $R$ in the calculation of derivatives used in the "downward" truncated Taylor series.
The fix to this problem appears to be simple since, for the region $1.8396 \leq x \leq 20$ and $y \leq 0.031623$, we may use "upward" truncated Taylor series (Taylor expansion about $z_0=x$) to evaluate the function. Only 7 terms from the "upward" Taylor series are required to attain 13 significant-digits accuracy in this region. However, the use of upward Taylor series expansion necessitates the computation of $w(x)$ which in turn depends on Dawson's integral for a real argument, $F(x)$, according to the relation

$$w(x) = e^{-x^2} + \frac{2i}{\sqrt{\pi}} F(x) \qquad (2),$$

The expansion coefficients are calculated recursively [Armstrong 1967, Shippony and Read 1993], where

$$d_0 = F(x), \quad d_1 = 1 - 2x d_0, \quad d_{n+1} = \frac{2}{n+1}(x d_n + d_{n-1}) \text{ for } n = 1, 2, ... \qquad (3)$$

Here we use Algorithm 715 [Cody 1993] to calculate Dawson's function of a real argument, $F(x)$, to the required accuracy.

These simple modifications were implemented in the *Fortran* code of Algorithm 680, tested using Algorithm 916 as a reference and were found to restore the accuracy to its required level for arguments near the real axis. In addition, the present correction significantly improves the efficiency of the code in the region of interest.

It has to be mentioned that Gautschi [Gautschi 1970] highlighted the problem of calculating $w(z)$ when $y$ is relatively small. He presented the possibility of using the Dawson integral and Taylor expansion to calculate $w(z)$; however, he rejected it based on the expectation that it would increase the execution time for computing $w(z)$ due to the necessity of computing $F(x)$ and that the recursive computation of the expansion coefficients is subject to considerable loss of accuracy, particularly for large $x>0$. The availability of Algorithm 715 [Cody 1993] to compute $F(x)$ efficiently and the restriction of the implementation of the correction to $x \leq 20$ remove Gauschi's concerns and clearly provides a suitable correction.

## 3. CONCLUSIONS

The cause of the loss-of-accuracy problem experienced when using Algorithm 680 to evaluate the Faddeyeva function near the real axis has been identified and a remedy for the problem is described and implemented.

## ACKNOWLEDGMENTS
The author would like to acknowledge helpful and insightful comments and suggestions received from the Algorithm Editor and an anonymous reviewer.